
\input harvmac
%
%
%
%
\ifx\answ\bigans
\else
\output={
  \almostshipout{\leftline{\vbox{\pagebody\makefootline}}}\advancepageno
}
\fi
%
%
%

%
%

%
%
\def\UCSD#1#2{\noindent#1\hfill #2%
\bigskip\supereject\global\hsize=\hsbody%
\footline={\hss\tenrm\folio\hss}}
%
%
\def\abstract#1{\centerline{\bf Abstract}\nobreak\medskip\nobreak\par #1}
%
%
%
%
\edef\tfontsize{ scaled\magstep3}
 \tfontsize  \tfontsize
 \tfontsize \font\titlei=cmmi10 \tfontsize
\font\titleis=cmmi7 \tfontsize \font\titleiss=cmmi5 \tfontsize
\font\titlesy=cmsy10 \tfontsize \font\titlesys=cmsy7 \tfontsize
\font\titlesyss=cmsy5 \tfontsize  \tfontsize
\skewchar\titlei='177 \skewchar\titleis='177 \skewchar\titleiss='177
\skewchar\titlesy='60 \skewchar\titlesys='60 \skewchar\titlesyss='60
%
%
%
%
%
\def\inv{^{\raise.15ex\hbox{${\scriptscriptstyle -}$}\kern-.05em 1}}
\def\lbar{{\lower.35ex\hbox{$\mathchar'26$}\mkern-10mu\lambda}} 

%
%
%
%
\def\dsl{\,\raise.15ex\hbox{/}\mkern-13.5mu D} 
\def\delsl{\raise.15ex\hbox{/}\kern-.57em\partial}
\def\Ksl{\hbox{/\kern-.6000em\rm K}}
\def\Asl{\hbox{/\kern-.6500em \rm A}}
\def\Dsl{\hbox{/\kern-.6000em\rm D}} 
\def\Qsl{\hbox{/\kern-.6000em\rm Q}}
\def\gradsl{\hbox{/\kern-.6500em$\nabla$}}
%
%
\def\lspace{\ifx\answ\bigans{}\else\qquad\fi}
\def\lbspace{\ifx\answ\bigans{}\else\hskip-.2in\fi} 
%
%
\def\boxeqn#1{\vcenter{\vbox{\hrule\hbox{\vrule\kern3pt\vbox{\kern3pt
        \hbox{${\displaystyle #1}$}\kern3pt}\kern3pt\vrule}\hrule}}}
%
%
\def\mbox#1#2{\vcenter{\hrule \hbox{\vrule height#2in
\kern#1in \vrule} \hrule}}
%
%
%
%

 \def\CJ{{\cal J}}  
  \def\CO{{\cal O}}

%
%
%
%
%

%

\def\bar#1{\overline{#1}}

\def\bra#1{\left\langle #1\right|}
\def\ket#1{\left| #1\right\rangle}

\def\darr#1{\raise1.5ex\hbox{$\leftrightarrow$}\mkern-16.5mu #1}

%
%
\def\frac#1#2{{\textstyle{#1\over #2}}} 
%
%
%
%

%
%
%
%

%
%
\def\ltap{\ \raise.3ex\hbox{$<$\kern-.75em\lower1ex\hbox{$\sim$}}\ }
\def\gtap{\ \raise.3ex\hbox{$>$\kern-.75em\lower1ex\hbox{$\sim$}}\ }
\def\gl{\ \raise.5ex\hbox{$>$}\kern-.8em\lower.5ex\hbox{$<$}\ }
\def\roughly#1{\raise.3ex\hbox{$#1$\kern-.75em\lower1ex\hbox{$\sim$}}}
%
%

%

%

\def\pl#1#2#3{{Phys. Lett. } B{#1} (#2) #3}

\relax

\noblackbox

\def\asl{\hbox{/\kern-.6500em A}}

\def\clebsch#1#2#3#4#5#6{\left(\left.
\matrix{#1&#2\cr#4&#5\cr}\right|\matrix{#3\cr#6}\right)}
\def\threej#1#2#3#4#5#6{\left\{
\matrix{#1&#2&#3\cr#4&#5&#6\cr}\right\} }
\vskip 1.in
\centerline{{\titlefont{Hyperfine Mass Splittings of Baryons}}}
\medskip
\centerline{{\titlefont{Containing a Heavy Quark in Large N QCD}}}
\vskip .3in
\centerline{Elizabeth Jenkins}
\vskip .2in
\centerline{\sl Department of Physics}
\centerline{\sl University of California, San Diego}
\centerline{\sl 9500 Gilman Drive}
\centerline{\sl La Jolla, CA 92093}
\vfill
\abstract{The hyperfine mass splittings of baryons containing a heavy
quark are derived at leading order in large $N$ QCD.  Hyperfine
splittings either preserve or violate heavy quark spin
symmetry.  Previous work proves that the splittings which
preserve heavy quark spin symmetry are proportional to ${\bf J}^2$ at
order $1/N$, where $J$ is the angular momentum of the light degrees of
freedom of the baryon.  This work proves that the splittings which
violate heavy quark spin symmetry are proportional to ${\bf J} \cdot
{\bf S_Q}$ at order $1/(N m_Q)$ in the $1/N$ and  $1/m_Q$
expansions.}
\vfill  \UCSD{\vbox{\hbox{UCSD/PTH 93-20}
\vskip-0.1truecm \hbox{hep-ph/9307245}}}{July 1993}   
\eject

The large $N$ expansion provides a quantitative and systematic method
for the calculation of baryonic properties in QCD.  Recent work
\ref\dm{R. Dashen and A.V. Manohar, UCSD/PTH 93-16}\ref\ejone{E.
Jenkins, UCSD/PTH 93-17}\ref\rdam{R. Dashen and A.V. Manohar,
UCSD/PTH 93-18}
proves that baryon-pion couplings are determined by a single
uncalculable coupling constant in large $N$.  The couplings derived
in large $N$ QCD satisfy light quark spin-flavor symmetry relations.
These symmetry relations are robust upto corrections of order
$1/N^2$ since the leading order $1/N$ correction vanishes.  Large $N$
predictions for other properties of baryons
also can be considered.  This work focuses on the hyperfine mass
splittings of baryons containing a single heavy quark.  The
splittings fall into two categories: splittings which preserve
heavy quark spin symmetry and splittings which violate heavy quark
spin symmetry.  Previous work \ref\j{E. Jenkins,
UCSD/PTH 93-19}\ proves that the spin symmetry-preserving splittings
are proportional to ${\bf J}^2$ at leading order in the $1/N$
expansion, where $J$ is the angular momentum of the light degrees
of freedom of the baryon.  This letter proves that the leading order
spin symmetry-violating splittings are proportional to ${\bf J}\cdot
{\bf S_Q}$.  Ref.~\dm\ proves that baryon mass splittings are first
allowed at order $1/N$ in the $1/N$ expansion.  The spin-symmetry
violating splittings are suppressed by an additional factor of $1/m_Q$
since heavy quark spin symmetry is a good symmetry in the infinite
heavy quark mass $m_Q \rightarrow \infty$ limit \ref\iswis{N. Isgur
and M.B. Wise, \pl {232} {1989}{113}, \pl {237}{1990}{527} }.  These
large $N$ results are identical to the predictions of the large $N$
Skyrme and non-relativistic quark models \ref\jmhyper{E. Jenkins and
A.V. Manohar, \pl {294}{1992}{273} }.

The proof of the above result begins with a discussion of the
spectrum of baryon states in large $N$.  In the $m_Q \rightarrow
\infty$ limit, heavy baryon states are constructed from the spin
states of the $S_Q = \frac 1 2$ heavy quark and the states
for the light degrees of freedom (which, by definition, include
everything except the heavy quark) of the baryon.  In large $N$ with
$N$ odd and $N_f = 2$ light flavors, the light degrees of freedom
consist of the degenerate tower of $(I,J)$ states $(0,0)$, $(1,1)$,
$(2,2)$, ...., $((N -1)/2, (N -1)/2)$, where $I$ and $J$ are the
isospin and the angular momentum of the light degrees of freedom,
respectively.  The $(0,0)$ state in the tower corresponds to a single
baryon, the spin-$\frac 1 2$ $\Lambda_Q$ baryon, with the spin of the
baryon determined by the spin of the heavy quark.  All other $(I,J)$
states correspond to a degenerate doublet of heavy baryon multiplets
with isospin $I$ and total spin equal to $I \pm \frac 1 2$, since
$J=I$ for the given tower of states.  For instance, the pair of baryon
states with light degrees of freedom given by the $(1,1)$ state
corresponds to the spin-$\frac 1 2$ $\Sigma_Q$ and the spin-$\frac 3 2$
$\Sigma_Q^*$.   All the heavy quark baryon states are degenerate upto
mass splittings of order $1/N$ \dm.  The $1/N$ mass splittings can be
divided into two categories.  Hyperfine splittings amongst states in
the tower produce baryon mass
splittings which do not depend on the heavy quark spin.  An example of
this type of splitting is the $(\Sigma_Q)_{\rm ave}-\Lambda_Q$ mass
difference, where $(\Sigma_Q)_{\rm ave}$ is the spin-averaged mass of
the $\Sigma_Q$ and $\Sigma_Q^*$ baryons.  The heavy quark
spin-independent mass splittings are identical to the hyperfine mass
splittings of the tower of states for the light degrees of freedom.
Ref.~\j\ proves that these splittings are proportional to ${\bf J}^2$
at leading order in large $N$.  This work concentrates on hyperfine
mass splittings which depend on the heavy quark spin.  For the
remainder of this paper, these splittings will be referred to as
heavy quark splittings in order to distinguish them from the heavy
quark spin-independent splittings.  An example of this type of
splitting is the $\Sigma_Q^* - \Sigma _Q$ mass difference.
An analogous heavy  quark splitting exists for each state in the
tower.  These heavy quark spin-dependent splittings arise at order
$1/m_Q$ because they violate heavy quark spin symmetry.  This paper
proves that the heavy quark spin-dependent mass splittings are
proportional to ${\bf J}\cdot {\bf S_Q}$ at leading order in large $N$
QCD.  The leading order splitting occurs at order $1/(N\,m_Q)$ in the
$1/N$ and $1/m_Q$ expansions.

The derivation of this result for the heavy quark
splittings is similar to the derivation for the hyperfine
splittings given in Ref.~\j.  For this
reason, many of the details of this calculation are not repeated
here.  The starting point of the proof is the consideration of chiral
logarithmic corrections to baryon masses due to pion loops.  For the
present calculation, the discussion given in Ref.~\j\ must be extended
to include separate renormalizations of the two baryon masses
corresponding to each $(I,J)$ state of the tower, such as the
$\Sigma_Q$ and $\Sigma_Q^*$ for the $(1,1)$ state.  The chiral
logarithmic correction to a baryon mass difference is equal to the
difference of the chiral logarithmic corrections to each of the
masses.  The crucial point of the proof is the realization that these
loop corrections grow with one power of $N$ more than a baryon mass
splitting.  Consistency of the large $N$ limit requires that the loop
correction to a baryon mass difference be the same order or higher
order in the $1/N$ expansion as the mass difference.  Thus,
consistency of the large $N$ limit requires an exact cancellation
amongst the chiral logarithms at leading order.  The condition of
exact cancellation results in   equations relating baryon mass
differences.  These equations have a unique solution, and they
determine all ratios of baryon mass differences.

The derivation of the large $N$ consistency conditions for the heavy
quark splittings is completely analogous to the
derivation of Ref.~\j.  The diagrams contributing to the chiral
logarithmic corrections are proportional to a single Feynman diagram,
so all kinematic factors factor out of the consistency conditions.
Evaluation of Clebsch-Gordan factors arising from each diagram is
considerably more involved in the case at hand, however, since the pion
couplings to heavy quark baryon states must be used rather than pion
couplings to the light degrees of freedom, which are simpler.  These
baryon-pion couplings are derived in Ref.~\ejone.  The couplings are
of the form
\eqn\bpib{
\bar B^\prime \,G^{ai} B \ {{\partial^a \pi^i} \over f_\pi}\ ,
}
where $a=1,2,3$ labels the angular momentum channel of the $p$-wave
pion,  $i=1,2,3$ labels the isospin of the pion, and $G^{ai}$ is an
operator with unit spin and isospin.  The matrix elements of
$G^{ai}$ for heavy baryon states $\ket{ I I_{z}, \CJ \CJ_{z} }$ with
isospin $I$, third component of isospin $I_z$, spin $\CJ$, and third
component of spin $\CJ_z$ are parametrized in terms of a
single unknown coupling constant in large $N$,
\eqn\hbcouplings{\eqalign{ &\bra{ I^\prime
I_{z}^\prime, \CJ^\prime \CJ_{z}^\prime } G^{ai}  \ket{ I I_{z}, \CJ
\CJ_{z} } = N \, g^\prime \,(-1)^{1+I+S_Q+\CJ^\prime}
\,(-1)^{I^\prime-I-1}\,(-1)^{\CJ^\prime-\CJ-1}\cr
&\qquad
\sqrt{ { (2I+1) } \,
{ (2\CJ+1) } } \threej{1}{I}{I^\prime}{S_Q}{\CJ^\prime}{\CJ}
\ \clebsch {I}{1}{I^\prime}{I_{z}}{i}{I^\prime_{z}}
\clebsch {\CJ}{1}{\CJ^\prime}{\CJ_{z}}{a}{\CJ^\prime_{z}}, \cr
}}
where $g^\prime$ is the arbitrary coupling constant of $\CO(1)$,
and the quantity in curly braces is the $6j$ symbol.  The
spin of the initial heavy quark baryon is either $\CJ=I + \frac
1 2$ or $\CJ = I - \frac 1 2$, whereas the spin of the final baryon
is either $\CJ^\prime= I^\prime + \frac 1 2$ or $\CJ^\prime =
I^\prime - \frac 1 2$, since $S_Q = \frac 1 2$ and the assumed tower
for the light degrees of freedom implies that the angular momentum of
the light degrees of freedom is equal to its isospin.  Since the
heavy quark carries zero isospin, the isospin of the light degrees of
freedom is also the isospin of the heavy quark baryon.  The
Clebsch-Gordan factors for the pion loop diagrams can be evaluated
using Eq.~\hbcouplings, the definition of the $6j$ symbol, and the
identity
\eqn\clbidentity{\eqalign{ \sum_{I_{2z},\, \CJ_{2z},\, i,\,
a} (-1)^i \, &(-1)^a  \clebsch{I_1}{1}{I_2}{I_{1z}}{i}{I_{2z}}
\clebsch{\CJ_1}{1}{\CJ_2}{\CJ_{1z}}{a}{\CJ_{2z}}\cr
&\quad\clebsch{I_2}{1}{I_1}{I_{2z}}{-i}{I_{1z}}
\clebsch{\CJ_2}{1}{\CJ_1}{\CJ_{2z}}{-a}{\CJ_{1z}}\cr &\qquad\qquad=
(-1)^{I_1-I_2}\,(-1)^{\CJ_1-\CJ_2} \sqrt{ { {(2I_2 + 1)} \over {(2 I_1
+1)} } { {(2\CJ_2 + 1)} \over {(2 \CJ_1 +1)} } }
\  ,\cr
}}
which sums over the $z$-components of isospin and spin of
intermediate baryon and pion states in a loop diagram.  A special
case of Eq.~\clbidentity\ is given in Ref.~\j.  After considerable
manipulation, a very simple consistency condition relating the
heavy quark splittings is obtained.

The consistency condition for the heavy quark splitting $\Delta M_i
\equiv B_i^* - B_i$ amongst baryons with light degrees of
freedom corresponding to the $i^{\rm th}$ state in the tower is given
by
\eqn\recursioni{ \Delta M_{i-1}\, { {J_{i-1}}
\over {J_{i}} } - \Delta M_{i} \left[{ {J_i} \over {J_{i+1}} }+
{ {J_{i+1}} \over {J_{i}} }\right]
+ \Delta M_{i+1}\, { {J_{i+2}} \over
{J_{i+1}} }=0 \ ,  }
where the angular momenta of the light degrees of freedom satisfy
$ J_{i+1} = J_i + 1 $
and $J_i = I_i$ for the assumed tower.  Thus, the $i=1$ state
corresponds to the $\Lambda_Q$ baryon with $J_1=0$, whereas the
$i=2$ state corresponds to the $\Sigma_Q$ and $\Sigma_Q^*$ baryons
with $J_2 =1$.  Note that the consistency condition for the
heavy quark splitting $\Delta M_i$ depends only on the splittings
$\Delta M_{i-1}$, $\Delta M_i$, and $\Delta M_{i+1}$ of the
$(i-1)$, $i$, and $(i+1)$ states in the tower because single pion
exchange can only change the spin or isospin of the initial baryon by
one unit.  Hence, the allowed intermediate baryon states in one-loop
diagrams for an external state with light degrees of freedom $i$ are
the states with light degrees of freedom $(i-1)$, $i$, and $(i+1)$.
The consistency condition Eq.~\recursioni\ for the heavy quark
splitting of state $i$ assumes that the splitting $\Delta M_{i-1}$
exists.  This is not the case for the first non-vanishing heavy quark
splitting $\Delta M_2=(\Sigma_Q^* - \Sigma_Q)$ since the first state
in the tower, $(0,0)$, corresponds to a single baryon state and hence
there is no analogue of $\Delta M$ for this state.   The $\Delta M_2$
heavy quark splitting satisfies a truncated version of
Eq.~\recursioni, given by    \eqn\recurinit{ - \Delta M_2 \left[{
{J_2} \over {J_3} }+ { {J_3} \over {J_2} }\right] + \Delta M_3\, {
{J_4} \over {J_3} }=0 \ ,
}
where $J_2 =1$.  The solution to Eqs.~\recursioni\
and~\recurinit\ is unique.  Given a hyperfine mass splitting $\Delta
M_2$, Eq.~\recurinit\ determines $\Delta M_3$.  Given $\Delta M_2$ and
$\Delta M_3$, Eq.~\recursioni\ then determines the mass splitting
$\Delta M_4$.  All remaining mass splittings are determined
recursively using Eq.~\recursioni.

The unique solution to the recursion relations~\recursioni\
and~\recurinit\ produces heavy quark splittings with the same
ratios as those produced by the operator ${\bf J}\cdot {\bf S_Q}$.  The
proof of this assertion is as follows.  Consider the initial recursion
relation Eq.~\recurinit.  This equation fixes the ratio $\Delta M_3/
\Delta M_2$,
\eqn\ratioone{
{ {\Delta M_3} \over {\Delta M_2} }
= { {J_2^2 + J_3^2 } \over {J_2 J_4} } \ ,
}
which must be compared with the ratio produced by the
operator ${\bf J}\cdot {\bf S_Q}$.  The difference of ${\bf J}\cdot
{\bf S_Q}$ values for states with total spin $J + \frac 1 2$ and
$J - \frac 1 2$ is given by
\eqn\jsdiff{
\frac 1 2 \left( (J + \frac 3 2) (J + \frac 1 2) - (J - \frac 1 2) (J +
\frac 1 2) \right)= \frac 1 2 (2J +1) \ ,
}
since ${\bf J}\cdot {\bf S_Q} = \frac 1 2 ( {\bf \CJ}^2 - {\bf J}^2 -
{\bf S_Q}^2 )$, and ${\bf J}^2$ and ${\bf S_Q}^2$ are the same for the
$\CJ = J + \frac 1 2$ and $\CJ = J - \frac 1 2$ states.  Thus, the
ratio obtained from ${\bf J}\cdot {\bf S_Q}$ is
\eqn\ratiotwo{ { {\Delta M_3} \over {\Delta
M_2} } ={ { (2J_3+1) } \over { (2J_2+1) } } \ .
}
Eqs.~\ratioone\ and~\ratiotwo\ are not equivalent for arbitrary
$J_2$.  However, for $J_2 = 1$, both expressions yield
\eqn\halfinit{
\Delta M_3 = \frac 5 3 \,\Delta M_2 \ .
}
It remains to prove that given an initial splitting ratio of $(2J_3
+1)/ (2J_2 +1)$, all subsequent ratios satisfy
\eqn\ratioj{
{ {\Delta M_{i+1}} \over {\Delta M_{i}} }= { {(2J_{i+1}+1)} \over
{(2J_i+1)} }  \ .
}
The recursion relation Eq.~\recursioni\ yields
\eqn\ratioi{
{ {\Delta M_{i+1}} \over {\Delta M_{i}} }=
{ {J_{i+1}} \over {J_{i+2}} }
\left\{
\left[  { {J_{i+1}} \over {J_{i}} } +
{ {J_{i}} \over {J_{i+1}} }   \right]
-  { {\Delta M_{i-1}} \over {\Delta M_{i}} }
\,{ {J_{i-1}} \over {J_{i}} }
\right\} \ .
}
Assuming that the ratio
\eqn\assum{
{ {\Delta M_{i}} \over {\Delta M_{i-1}} }
= { {(2J_{i}+1)} \over {(2J_{i-1}+1)} } \ ,
}
Eq.~\ratioi\ then implies that Eq.~\ratioj\ is satisfied.  This
result is most easily seen by making the substitution $J_i = \frac n
2$.  Then the sought result is equivalent to the identity
\eqn\niden{
{ {(n+2)} \over {(n+4)} }
\left\{
\left[
{ {(n+2)} \over {n} } + { {n} \over {(n+2)} }
\right]
 - { {n-1} \over {(n+1)} } { {(n-2)} \over {n} }
\right\}
= { {(n+3)} \over {(n+1)} } \ ,
}
which is true for arbitrary $n$.  This completes the proof that the
heavy quark splittings are proportional to ${\bf J}\cdot {\bf S_Q}$.

It is worth noting that the heavy quark splittings $\Delta M_i$ are in
the same ratios as the nucleon hyperfine mass splittings derived in
Ref.~\j.  There is a simple reason for this unanticipated
result.  The recursion relation derived for the ${\bf J}\cdot {\bf
S_Q}$ heavy quark splittings is very similar to that
for the ${\bf J}^2$ hyperfine splittings.  The recursion relation
derived in Ref.~\j\ is of the form
\eqn\xy{
{1 \over X_{i-1}}{1 \over Y_{i-1}}
-\left[ {1 \over X_i} + X_i \right]
+ X_{i+1} Y_{i+1} =0 \ ,
}
where
\eqn\xi{
X_i = { {2 J_{i+1} +1} \over {2 J_{i} +1} }=
{ {J_{i+1}+ \frac 1 2} \over {J_i+ \frac 1 2} }
}
are the coefficients of the recursion equation and
\eqn\yi{
Y_i = { {J_{i+1} } \over {J_{i}} }
}
are equal to the ratios of
hyperfine mass splittings,
\eqn\ym{
Y_i = { {(M_{i+1} - M_i)} \over {(M_i - M_{i-1})} } \ .
}
For the nucleon hyperfine mass splittings, such as the
splitting $\Delta - N$, the $J_i$ take half-integral values.  Note that
the $X_i$ for half-integral values of $J_i$ are equal to the $Y_i$ for
integral values of $J_i$, with $X_1(J_1 =1/2)=Y_2(J_2 =1)$.  Thus,
Eq.~\xy\ for half-integral $J_i$ is identical to the recursion
relation of the heavy quark splittings for integral $J_i$,
\eqn\yx{ {1 \over
X_{i-1}}{1 \over Y_{i-1}} -\left[ {1 \over Y_i} + Y_i \right]
+ X_{i+1} Y_{i+1} =0 \ ,
}
where the $Y_i$ are the coefficients of the recursion equation and
the $X_i$ are the ratios of heavy quark splittings,
\eqn\xm{
X_i = { {\Delta M_{i+1}}\over {\Delta M_i} }\ .
}

The derivation of this paper assumed that the number of colors $N$ was
odd.  The results obtained for $N$ odd are also true for $N$ even.
For an even number of colors, the light degrees of freedom of the heavy
quark baryons form the tower of half-integral $(I, J)$ states $(1/2,
1/2)$, $(3/2, 3/2)$, $(5/2, 5/2)$, ..., $((N-1)/2, (N-1)/2)$.  In this
case, the first non-vanishing heavy quark splitting is the splitting
$\Delta M_1$ corresponding to the first state in the tower with $J_1 =
\frac 1 2$.  The consistency condition for $\Delta M_1$
implies
\eqn\zeroinit{ \Delta M_2 = 2 \,\Delta M_1 \ . }
The recursion relation for the heavy quark splitting of state $i$ is
identical to Eq.~\recursioni.  The unique solution of
Eqs.~\recursioni\ and~\zeroinit\ yields heavy quark splittings
proportional to ${\bf J}\cdot {\bf S_Q}$.  When the number of
colors $N$ is even, the spectrum of baryons which contain no heavy
quarks is given by the integral $(I,J)$ tower.  Consideration of
Eqs.~\xy\ and \yx\ shows that the ratios of heavy quark
splittings for half-integral angular momentum of the light degrees of
freedom are identical to the ratios of hyperfine splittings for the
integral $(I,J)$ tower of baryon states.  In this case $X_1(J_1=0) =
Y_1(J_1=1/2)$.

In conclusion, this work proves that the heavy quark splittings of
baryons containing a single heavy quark are proportional to ${\bf
J}\cdot {\bf S_Q}$.  The splittings also are proportional to
$1/(N\,m_Q)$, since consistency of the large $N$ expansion requires
degeneracy of the baryon states upto order $1/N$ \dm, and heavy quark
spin symmetry violation first occurs at order $1/ m_Q$.

\vfill\break\eject

\centerline{\bf Acknowledgements}
I thank the Aspen Center for Physics
for hospitality while this work was completed.
This work was supported in part by the Department of Energy
under grant number DOE-FG03-90ER40546.

\listrefs
\vfill
\eject

\bye